\documentclass[prb,aps,twocolumn,showpacs,amsmath,amssymb,floatfix,tighten,bibtex]{revtex4-1}
\usepackage{graphicx,amsmath,amsfonts,latexsym,bm,eso-pic}
\usepackage{graphicx}
\usepackage{bm}
\usepackage{amssymb}
\usepackage{color}
\usepackage{epsfig}
\usepackage{subfigure}

\newcommand{\be}{\begin{equation}}
\newcommand{\ee}{\end{equation}}
\def\eq{&=&} 
\def\nn{\nonumber}

\def\der#1#2{\frac{\partial #1}{\partial #2}}

\def \be{\begin{equation}}
\def \ee{\end{equation}}
\def \ba{\begin{array}}
\def \ea{\end{array}}
\def \bea{\begin{eqnarray}}
\def \eea{\end{eqnarray}}
\def \nn{\nonumber}
\def \l{\left}

\def \rr{\right}

\def \H{{\cal{H}}}

\def\to{\tilde{\omega}}
\def\td{\tilde{\Delta}}

\def\sx{\sigma_x}
\def\sy{\sigma_y}
\def\sz{\sigma_z}

\def\tz{\tau_z}
\begin{document}

\title {Disordered topological metals}

\author{Julia S. Meyer$^1$, Gil Refael$^{2,3}$}
\affiliation{$^1$ SPSMS, UMR-E 9001 CEA/UJF-Grenoble 1, INAC, Grenoble, F-38054, France\\
$^2$ Dept. of Physics, California Institute of Technology, MC 149-33, 1200 E. California Blvd., Pasadena, CA, 91125, USA\\
$^3$ Dahlem Center for Complex Quantum Systems and Fachbereich Physik, Freie Universit\"at Berlin, 14195 Berlin, Germany}

\begin{abstract}
Topological behavior can be masked when disorder is present. A
topological insulator, either intrinsic or interaction induced, may
turn gapless when sufficiently disordered. Nevertheless, the metallic
phase that emerges once a topological gap closes retains several
topological characteristics. By considering the self-consistent
disorder-averaged Green function of a topological insulator, we derive
the condition for gaplessness. We show that the edge states survive
in the gapless phase as edge resonances and that, similar to a doped
topological insulator, the disordered topological metal also has a
finite, but non-quantized topological index. We then consider the
disordered Mott topological insulator. We show
that within mean-field theory, the disordered Mott topological
insulator admits a phase where the symmetry-breaking order parameter
remains non-zero but the gap is closed, in complete analogy to \lq
gapless superconductivity\rq\ due to magnetic disorder.

\end{abstract}

 \maketitle

\section{Introduction} 

The discovery of topological insulators (TI) has brought us to nothing
less than a revision of electronic band structure theory. Topological
behavior was predicted for two-dimensional (2d) systems such as
graphene with spin-orbit coupling \cite{KaneMele1,KaneMele2} and HgTe
quantum wells \cite{BHZ}. While the spin-orbit coupling in graphene
turns out to be too weak, topological behavior has indeed been
observed in HgTe/CdTe heterostructures \cite{Molenkamp}. Shortly
thereafter, three-dimensional (3d) topological insulators were
predicted to exist in bismuth alloys
\cite{MooreBalents,FuKane3D,FuKaneMele3D}, and by now have been
discovered in a variety of materials (see, e.g.,
Refs. \onlinecite{HasanTernery,HasanFam}). From a theoretical
perspective, these materials are described by a non-zero quantized
topological index, which is essentially a generalization (albeit a
nontrivial one) of the Chern number
\cite{Volovik,MooreBalents,FuKaneMele3D,FuKane3D,QiZhangTR,QiZhangindex}. As
in Hall insulators, the topological index implies the existence of
protected edge/surface states at the boundaries of the system
\cite{Volovik,Gurarie}, which are responsible for a variety of
exciting aspects of topological insulators. In two dimensions, the 1d
edge states yield a quantized two-terminal conductance; in three
dimensions, the 2d surface states consist of single helical Dirac
cones, and gapping them leads to an anomalous surface Hall response
and the so-called axion magneto-electric response
\cite{QiZhangTR}. The edge and surface states are at the heart of any
possible technological application for topological insulators,
including proposals to use them as a platform for Majorana states and
quantum computation
\cite{FuKaneMajorana,FranzMajorana,FuKaneMajorana2}.

Alongside ideal TIs, recently, the study of the effects of
imperfections -- which plague essentially all realizations so far  --
has also picked up. Experimentally, already the 2d heterostructure
realizations show the lack of quantized transport for samples of
intermediate length \cite{Molenkamp}. Furthermore, 3d topological
materials are often doped away from the bandgap, exhibiting a rather
large bulk conductance \cite{Paglione,Jarillo-Herrero}. Theoretical
studies of doping showed that the metal obtained when the chemical
potential of the TI lies outside the bandgap is characterized by a
finite, but non-quantized topological index \cite{BarkeshliQi,Bergman,
  BergmanRefael,HastingsLoring1,HastingsLoring2}. Studies of disorder
led to additional surprises: it was shown that disorder can induce
topological behavior in some trivial insulators
\cite{Jain,Beenakker,GuoFranzRefael}, and (more importantly for the
current work) that disorder may close the gap in topological
insulators, leading to a metallic phase
\cite{Shindou1,Shindou2,Prodan2D,Prodan3D}.

In our work, we focus on the disorder-induced metallic phase 
in a simple test case of the Kane-Mele-Haldane honeycomb model
\cite{Haldane,KaneMele1}. Naively, once the disorder destroys the gap
of a topological insulator, a simple metal emerges. By first
constructing the self-consistent disorder-averaged Green function, we
investigate the disorder-averaged topological index as well as the
fate of the edge states in the metallic phase. Our results indicate
that, on both counts, the disorder-induced metal retains some
signatures of its topological origin. Just like a doped TI, it has a
non-quantized, but finite topological index. Correspondingly, we show
that the edge states also survive as resonances of the
disorder-averaged Green function. In particular, the edges states
maintain their helical nature over a finite lifetime/mean free path,
and then get absorbed into the bulk. Surprisingly, their velocity is
strongly renormalized by the disorder, and vanishes at the transition
between the insulating and metallic phase.

As pointed out by several groups, next-nearest neighbor interactions
in various lattices may induce a topological phase
\cite{Raghu,Franz,Vishwanath,PesinBalents,YBKim}, namely a so-called
Mott topological insulator. Using our formalism, we self-consistently
analyze the formation of such a phase in the honeycomb lattice. We
show that, in close analogy to superconductivity in the presence of
magnetic impurities \cite{Gorkov,DeGennes,Maki}, disorder separates
the appearance of topological order into two transitions: one
associated with the appearance of a broken symmetry and an order
parameter, and one (at larger interactions or weaker disorder)
associated with the appearance of a spectral gap. The phase with a
finite order parameter but no gap coincides with the topological metal
we discussed before.

Below we start by developing the self-consistent disorder-averaged
formalism (Sec. \ref{GF}), and use it to find the critical disorder
strength for closing the TI's band gap (Sec. \ref{GTI}). We then
proceed to analyze the topological index, following Volovik's
prescription (Sec. \ref{topoind}). Perhaps the most physical
consequences of our discussion regard the edge states. In
Sec. \ref{edges} we derive the properties of the edge states, assuming
they correspond to a pole or resonance of the bulk disorder-averaged
Green function when an edge is introduced. Before concluding, we
discuss the possibility of an interaction-induced gapless topological
phase in the disordered honeycomb lattice, and identify the parameter
regime where this occurs (Sec. \ref{GMTI}).

\section{Disorder-averaged Green function \label{GF}}

We consider two copies of the Haldane Hamiltonian for a single-spin
anomalous Hall state \cite{Haldane} or, equivalently, the Kane-Mele
model for spin-1/2 fermions \cite{KaneMele1}. As we are interested in
the low-energy physics, we restrict ourselves to  the Hamiltonian near
its Dirac nodes,
\be
\H=v_Fk_x\sx\tz-v_Fk_y\sy-\Delta \sz s_z\tz.
\label{hh}
\ee
Here $\sigma_{\alpha}$ are Pauli matrices acting in pseudo-spin
(sublattice) space, while $\tau_z$ and $s_z$ are Pauli matrices
operating in valley and spin  space, respectively.

Our goal is to understand the properties of this model in the presence
of disorder which, if sufficiently strong, destroys its gap. Impurity
scattering is described by adding 
\begin{eqnarray}
{\cal H}_{\rm imp}=\sum_{i,j,k}U_{ijk}({\bf k}-{\bf k'})\sigma_is_j\tau_k,
\label{AG}
\end{eqnarray}
with $\langle U_{ijk}({\bf r})\rangle = 0$ and
 \begin{equation}
\langle U_{ijk}({\bf r}) U_{lmn}({\bf r'})\rangle = \gamma_{ijk} \,\delta({\bf r}-{\bf r'})\delta_{il}\delta_{jm}\delta_{kn},
\end{equation}
to the Hamiltonian (\ref{hh}). Here $\langle\dots\rangle$ denotes
disorder averaging.

Within the Born approximation, the self-energy due to impurity scattering reads 
\begin{eqnarray}
\Sigma(\omega)\eq\sum_{i,j,k}\gamma_{ijk}\,\sigma_is_j\tau_k\int (dk)\;G_0(\omega,{\bf k})\,\sigma_is_j\tau_k,
\end{eqnarray}
where
\begin{equation}
G_0(\omega,{\bf k})=\left(i\omega-v_F(k_x\sigma_x\tau_z-k_y\sigma_y)+\Delta\sigma_z s_z\tau_z\right)^{-1}
\end{equation}
is the unperturbed imaginary-time Green function of the system and $(dk)=dk_x\, dk_y/(2\pi)^2$. Upon momentum integration, one obtains
\begin{eqnarray}
\Sigma(\omega)\eq-\frac1{4\pi v_F^2}\ln\frac{D^2}{\omega^2+\Delta^2}\times\\
&&\times\sum_{i,j} \gamma_{ijk}\left[i\omega-\Delta(\sigma_i\sigma_z\sigma_i)(s_js_zs_j)(\tau_k\tau_z\tau_k)\right],\nn
\label{sigma}
\end{eqnarray}
where $D$ is a high-energy cut-off of order of the bandwidth.

One may distinguish two types of disorder,
$(\sigma_i\sigma_z\sigma_i)(s_js_zs_j)(\tau_k\tau_z\tau_k)=\pm
\sigma_z s_z\tau_z$. The strongest effect on the gap is due  to
impurities that obey
$(\sigma_i\sigma_z\sigma_i)(s_js_zs_j)(\tau_k\tau_z\tau_k)=\sigma_zs_z\tau_z$
which is the case, e.g., for relatively smooth potential disorder,
$i=j=k=0$. Therefore, in the following, we will restrict our attention
to this case.

To obtain the effective disorder-averaged Green function for this
problem, we note that the self-energy renormalizes
$\omega\rightarrow\tilde\omega$ and
$\Delta\rightarrow\tilde\Delta$. The self-consistent Born
approximation is obtained by substituting the renormalized $\to$ and
$\td$ in Eq.~\eqref{sigma}. The full Green function,
$G^{-1}=G_0^{-1}-\Sigma$, can then be written in the form
\begin{equation}
G(\omega,{\bf k})=\left(i\tilde\omega-v_F(k_x\sigma_x\tau_z-k_y\sigma_y)+\tilde\Delta\sigma_z s_z\tau_z\right)^{-1}\!,\!\!
\label{sc1}
\end{equation}
where
\begin{subequations}
\begin{eqnarray}
\omega\eq\tilde\omega(1-\zeta\ln\frac{D^2}{\tilde\omega^2+\tilde\Delta^2}),\label{eq-otilde}\\
\Delta\eq\tilde\Delta(1+\zeta\ln\frac{D^2}{\tilde\omega^2+\tilde\Delta^2}),
\end{eqnarray}
\label{sc2}
\end{subequations}
with $\zeta=\gamma_{000}/({4\pi v_F^2})$, using the definition of $\gamma_{ijk}$ in Eq. (\ref{AG}). 

In the absence of a topological insulator gap, $\Delta=0$, the above
analysis readily yields the mean-free time. At $\td=\Delta=0$,
analytic continuation of Eq. \eqref{eq-otilde} to real time yields
\be
\omega=\tilde\omega\left(1-\zeta\ln\frac{D^2}{(i\tilde\omega)^2}\right).
\ee
For $\omega=0$, we then obtain the mean free time \cite{Mirlin} at the
Dirac point, $\tau=i/\to(\omega=0)$. Namely,
\be
{\tau}^{-1}= D e^{-\frac{1}{2\zeta}}. 
\label{mft}
\ee

\section{Gapless topological phase \label{GTI}}

As noted by several authors
\cite{Shindou1,Shindou2,HastingsLoring1,HastingsLoring2,Prodan2D,Prodan3D},
disorder produces a phase transition between the topological insulator
phase and a gapless metallic phase. This can be seen most directly
using the self-consistent Green function, given by Eqs. (\ref{sc1})
and (\ref{sc2}). In fact, the problem of finding the renormalized gap
of the TI is equivalent to finding the renormalized gap of a
superconductor with pair-breaking disorder. As explained in a review
paper by Maki \cite{Maki}, upon analytic continuation to real
frequencies, the renormalized gap may be identified as  the largest
$\omega$ which permits a real solution for the renormalized $\to$ and
$\td$. This follows from the expression for the density of states,
\begin{eqnarray}
\nu(\omega)\eq-\frac1{8\pi}\textrm{Im}\left[\int (dk)\;{\rm Tr}\left[G(\omega,{\bf k})\right]_{i\omega\rightarrow\omega+i0^+}\right],
\end{eqnarray}
which upon using Eqs. \eqref{sc1} and \eqref{eq-otilde} yields
\begin{eqnarray}
\nu(\omega)=\frac1{8v_F^2\zeta}\textrm{Im}\left[\tilde\omega\right].
\end{eqnarray}
Performing the analytic continuation and rescaling all energies by $\Delta$,   Eqs. \eqref{sc2} simplify to
\begin{subequations}
\label{tstuff}
\begin{eqnarray}
\frac{\omega}{\Delta}\eq\nu\l(1-\zeta\ln
\frac{{\Lambda}^2}{\delta^2\!-\!\nu^2}\rr),\\
\delta^{-1}\eq1+\zeta\ln
\frac{{\Lambda}^2}{\delta^2\!-\!\nu^2},\label{eq-delta}
\end{eqnarray}
\end{subequations}
where $\delta=\td/\Delta$, $\nu=\to/\Delta$, and $\Lambda=D/\Delta$.
Combining the two equations, we can eliminate the logarithm to find
\be
\frac{\omega}{\Delta}=\nu\l(2-\frac{1}{\delta}\rr).\label{eq-om2}
\ee
Solving Eq. \eqref{eq-delta} for $\nu$ and inserting the solution
into Eq. \eqref{eq-om2}, we then obtain
\be
\frac{\omega}{\Delta}=\l(2-\frac{1}{\delta}\rr)\sqrt{\delta^2-{\Lambda}^2\exp\l[-\frac{1}{\zeta}\l(\frac{1}{\delta}-1\rr)\rr]},\label{eq-omegadelta}
\ee
which restricts the values of $\delta$ yielding $\omega\in{\mathbb
  R}_+$. On the one hand,
\be
\delta>\delta_{\rm min}=\frac{1}{2},
\ee
and, on the other hand, in order for the square root to be real, 
\begin{eqnarray}
\delta<\delta_{\rm max}=\Lambda \exp\l[-\frac{1}{2\zeta}\l(\frac{1}{\delta_{\rm max}}-1\rr)\rr].
\end{eqnarray}
The gap is then a function of $\zeta$ and given by the maximum of
$\omega$ for the range $\delta_{\rm min}\le\delta\le\delta_{\rm
  max}$. 

At a critical value of the disorder parameter $\zeta$, the gap
vanishes, and an insulator-metal transition occurs. This happens when
$\delta_{\rm max}=\delta_{\rm min}=1/2$ and the two zeros of $\omega$
as a function of $\delta$ merge. One obtains
\begin{eqnarray}
\frac12=\Lambda e^{-\frac1{2\zeta_{\rm gap}}(2-1)}\quad\Rightarrow\quad\zeta_{\rm gap}=\frac1{2\ln(2\Lambda)},\label{eq-zg}
\end{eqnarray}
or, equivalently,
\be
{\tau}_{\rm gap}^{-1}=\frac\Delta2
\ee
with $\tau$ given by Eq.~\eqref{mft}, i.e., $\tau_{\rm
  gap}^{-1}=D\exp[-1/(2\zeta_{\rm gap})]$. 

\section{Topological index \label{topoind}}

\begin{figure}[h]
\includegraphics[width=0.95\linewidth]{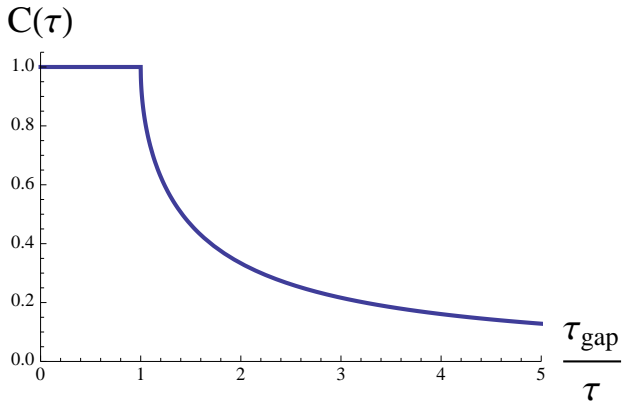}
\caption{The topological index for $s_z=-1$ as a function of the scattering rate
  $1/\tau$. As soon as
  $1/\tau$ exceeds $1/\tau_{\rm gap}=\Delta/2$, the topological index
  plunges from a quantized value to a nonquantized and monotonically
  decreasing, disorder-dependent value. }\label{fig-topinv}
\end{figure}

We are interested in determining whether the system, though gapless,
retains some topological properties. Let us start by considering the
disorder-averaged topological index of the system. 
There are several definitions for the topological index in 2d
\cite{Volovik,Qi-Zhang,QiZhangindex,QiZhangTR,FuKane3D}. For our
purposes, the most useful approach is the one of
Volovik~\cite{Volovik,QiZhangTR,Qi-Zhang} which calculates the Berry
phase directly from the Green function. The expression for the
topological index, see e.g. Eq.~(48) in Ref.~\onlinecite{Qi-Zhang},
for one spin species is given by
\begin{widetext}
\begin{eqnarray}
C\eq\frac1{2}\int d\omega\int (dk)\;\epsilon^{ij}{\rm Tr}\left[G\partial_\omega G^{-1}G\partial_{k_i}G^{-1}G\partial_{k_j}G^{-1}\right]_{\sigma,\tau}.\label{vindex}
\end{eqnarray}
\end{widetext}
In order to obtain the disorder-averaged topological index, we simply
substitute the disorder-averaged Green functions \cite{ft1}.
Using Eq. \eqref{sc1}, one obtains
\begin{eqnarray}
G\partial_\omega G^{-1}\eq G\left(i\partial_\omega\tilde\omega+\partial_\omega\tilde\Delta\sigma_zs_z\tau_z\right)\!,\;\;\\
\epsilon^{ij}G\partial_{k_i}G^{-1}G\partial_{k_j}G^{-1}\eq iv_F^2G\sigma_y\left\{\sigma_z\tau_z,G\right\}\sigma_y.
\end{eqnarray}
Evaluating the $k$-integrals and the trace, the expression can be reduced to 
\begin{eqnarray}
C\eq\frac{1}{\pi}s_z\int d\omega\;\left(\tilde\omega\partial_\omega\tilde\Delta-\tilde\Delta\partial_\omega\tilde\omega\right)\frac1{\tilde\omega^2+\tilde\Delta^2}.
\end{eqnarray}
Thus, one obtains
\begin{eqnarray}
C\eq-\frac{2}{\pi}s_z\arctan\frac1{x_{\rm min}},
\end{eqnarray}
where 
\begin{eqnarray}
x_{\rm min}\eq\frac{\tilde\omega(\omega=0)}{\tilde\Delta(\omega=0)}=\textrm{Re}\left[\sqrt{4\frac{D^2}{\Delta^2}e^{-\frac1\zeta}-1}\,\right],
\end{eqnarray}
or, using the mean free time,
\begin{eqnarray}
x_{\rm min}\eq\theta\left(1-\frac\Delta2\tau\right)\sqrt{\left(\frac\Delta2\tau\right)^{-2}-1}.
\end{eqnarray}
In the gapped region, $\tau^{-1}<\tau_{\rm gap}^{-1}=\Delta/2$, one
obtains $x_{\rm min}=0$ and therefore the topological constant is
$C=-s_z$. By contrast, if $\Delta=0$, one obtains $x_{\rm min}=\infty$
and therefore the topological constant vanishes, $C=0$. The gapless
region at $\tau^{-1}>\tau_{\rm gap}^{-1}$, however, emerges as the
most interesting. In this region $C$ changes smoothly from $-s_z$ to
$0$ (see Fig. \ref{fig-topinv}). Close to the closing of the gap,
$0<\tau^{-1}-\tau_{\rm gap}^{-1}\ll\tau_{\rm gap}^{-1}$, we find
\be
\delta C=C+s_z\propto\sqrt{\tau_{\rm gap}/\tau-1},
\ee 
whereas deep in the gapless phase, $\tau^{-1}\gg\tau_{\rm gap}^{-1}$ we find that the index falls off as 
\be
C\propto\tau/\tau_{\rm gap}.
\ee

A non-quantized topological index requires a bit of consideration. For
a given realization, by definition, Eq. (\ref{vindex}) yields a
quantized value. However, in the gapless phase, this value would
fluctuate between 0 and $-s_z$, depending on the specific disorder
realization. Upon averaging, one may, thus, obtain a non-quantized
value. It has already been pointed out by Loring and Hastings
\cite{HastingsLoring1,HastingsLoring2}, that the average topological
index of doped 2d TIs interpolates continuously between the respective quantized
values in the topologically non-trivial and trivial phases as the
chemical potential is varied and the valence band becomes partially
filled.  Non-quantized values have also been obtained in clean 3d
topological insulators, when the chemical potential is within the
valence or conduction band \cite{BarkeshliQi,Bergman}. 

The calculation we presented above fits well with the idea of a
partial response of a metal. It shows that the disorder-induced
gapless state should be quite similar in its behavior to a doped
topological insulator, with the chemical potential lying within one of
the bands. The index itself, however, is not a measurable
quantity. Below, we try to understand the physical ramifications of
this partial topological behavior in terms of the edge state physics.

\section{The fate of the edge state \label{edges}}

In the search for topological properties of the disordered metal,
several works direct us to the edge states. Recently
Ref. \onlinecite{Gurarie}~confirmed rigorously that a spatial jump of
the topological index (defined by Eq. \eqref{vindex}) results in an
edge state, i.e., a pole \cite{ft2} of the Green function at the
location of the jump. It is natural to ask what a non-quantized jump
implies for such edge states? 

Indeed, given the connections above, we speculate that the
non-quantized topological index implies an edge state resonance of the
Green function, but with a finite lifetime due to hybridization with
the continuum. 

To explore the edge physics, we start with the inverse of the disorder-averaged
Green function Eq. (\ref{sc1}) which may be associated with an
effective Schr\"odinger equation. For the four-component wave function
$\psi_{s_z}$ of a given spin species, the effective Schr\"odinger
equation reads
\begin{equation}
\left(v_Fk_x\sigma_x\tau_z-v_Fk_y\sigma_y-\tilde\Delta\sigma_z s_z\tau_z\right)\psi_{s_z}=\tilde\omega\psi_{s_z}.
\label{hedge}
\end{equation}
Consider now that the system occupies the half-plane $x<0$. To satisfy
the boundary conditions, we have to find solutions of \eqref{hedge}
that vanish on the edge, $x=0$.
We recall that the degree of freedom $\tau_z=\pm 1$  indicates the valley, i.e.,
the momentum around which $k_x,\,k_y$ in the Hamiltonian (\ref{hedge})
are measured. The true momenta are $p_x=k_x+\tau_zK_0$ and $p_y=k_y$,
where $K_0=4\pi/(3\sqrt3 a)$ and $a$ is the lattice spacing. For the
edge state to obey the boundary conditions, it is necessary to mix
wave functions from the two valleys. Namely, the generic form of the
edge-state wave function, assuming it is  given by a simple
superposition of momentum states (albeit with complex momenta), is
$\psi=\sum_{\tau_z=\pm1}\psi_{\tau_z}(x,y)e^{i\tau_zK_0x}$. This
implies that, for the wave function to vanish at the edge, both
components $\tau_z=\pm1$ of the wave function must be described by
identical sublattice ($\sigma$-space) spinors. 

To find such solutions, we separate the Schr\"odinger equation into
its parts dependent and independent of $\tau_z$, and require that both
vanish. Namely,
\begin{subequations}
\begin{eqnarray}
\left(v_Fk_x\sigma_x-\tilde\Delta\sigma_z s_z\right)\phi_{s_z}\eq0,\\
\left(\tilde\omega+v_Fk_y\sigma_y\right)\phi_{s_z}\eq0,
\end{eqnarray}
\end{subequations}
where $\phi_{s_z}$ is a two-component wave function in sublattice space.
In order for these equations to be solvable simultaneously, $k_x$ and
$k_y$ must obey
\begin{eqnarray}
k_x=\pm is_z\frac{\tilde\Delta}{v_F} \qquad{\rm and} \qquad k_y=\mp\frac{\tilde\omega}{v_F},\label{kxky}
\end{eqnarray}
while  the corresponding wave functions take the form
$\phi_{s_z}=(1,\pm i)^T/\sqrt 2$.

\begin{figure}[h]
\includegraphics[width=0.95\linewidth]{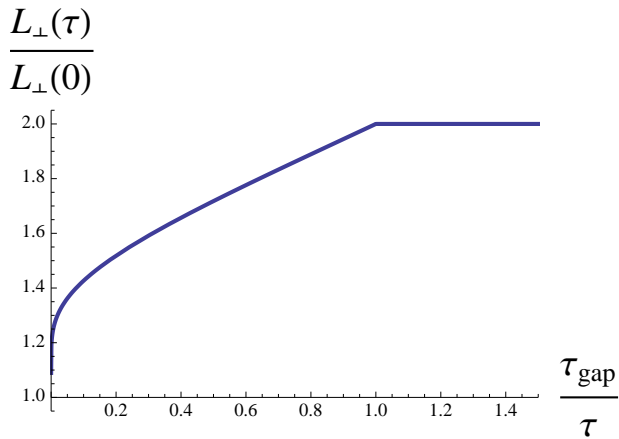}
\caption{Penetration of the edge state into the bulk: The transverse
  localization length of the edge state,
  $L_\perp(\tau)=v_F/\textrm{Re}[\tilde\Delta]$, in units of the bare
  decay length, $L_\perp(0)=v_F/\Delta$, is plotted as a
  function of disorder. It varies by a factor of
  $2$. }\label{fig-edge_perp}
\end{figure}

The complex valued momentum $k_x$ describes the decay of the edge
state away from the edge.  The decay length scale is, therefore, given
by $L_\perp=v_F/\textrm{Re}[\tilde\Delta]$. The sign has to be chosen
such that $\textrm{Im}[k_x]<0$. As can be seen in
Fig. \ref{fig-edge_perp}, the state remains well localized on the
edge. Namely, the transverse localization length only increases from
$L_\perp=v_F/\Delta$ at $\zeta=0$ to $L_\perp=2v_F/\Delta$ in the
gapless phase.

The momentum $k_y$ describes the propagation along the edge. The
appropriate choice of sign for $k_x$ yields
$k_y=-s_z\tilde\omega/v_F$. Thus, as expected,  the sign of $k_y$ and
therefore the propagation direction is tied to the spin. To obtain the
dispersion of the edge states, as well as their lifetime, we need to
find the relation between $k_y$ and the real energy $\omega$. We use
Eqs. \eqref{sc2} to write
\be
\omega=\tilde\omega\left(1-\zeta\ln\frac{D^2(2\tilde\omega-\omega)^2}{\tilde\omega^2(\Delta^2-(2\tilde\omega-\omega)^2)}\right).\label{om}
\ee
Since the edge state is helical, an imaginary part of the wave number
$k_y$ can be interpreted as a finite lifetime of the state. To
illustrate this, consider the simplest Schr\"odinger equation for a
decaying helical mode:
$i\der{\psi}{t}=-v\frac{1}{i}\der{\psi}{x}-i\frac{\psi}{\tau}$. This
is equivalently solved either by $\psi=\exp[i\omega (t-x/v)-t/\tau]$
or $\psi=\exp[i\omega (t-x/v)-x/(v\tau)]$. Which of these two
solutions should be used depends on the specific situation, and, for a
Green-function description, the two should be equivalent. 

\begin{figure}[h]
(a)\\
\includegraphics[width=0.95\linewidth]{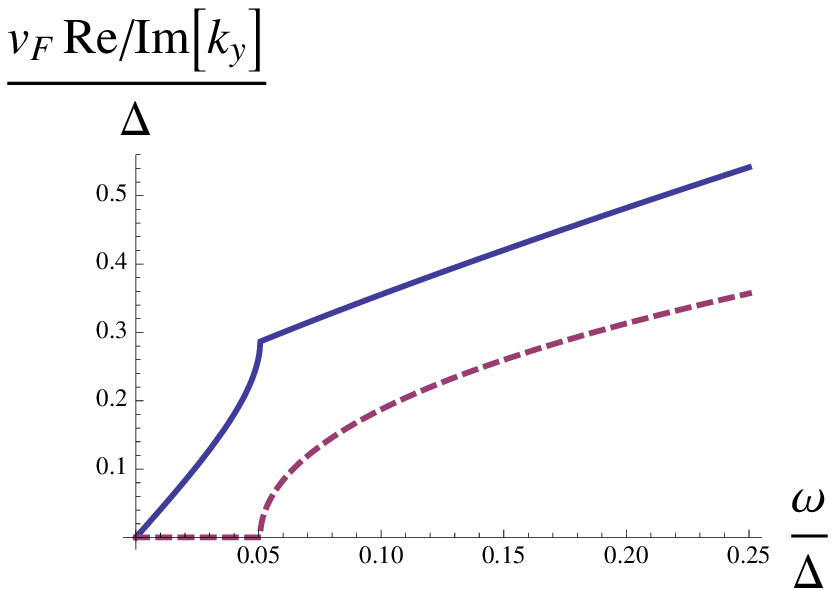}
\\
(b)\\
\includegraphics[width=0.95\linewidth]{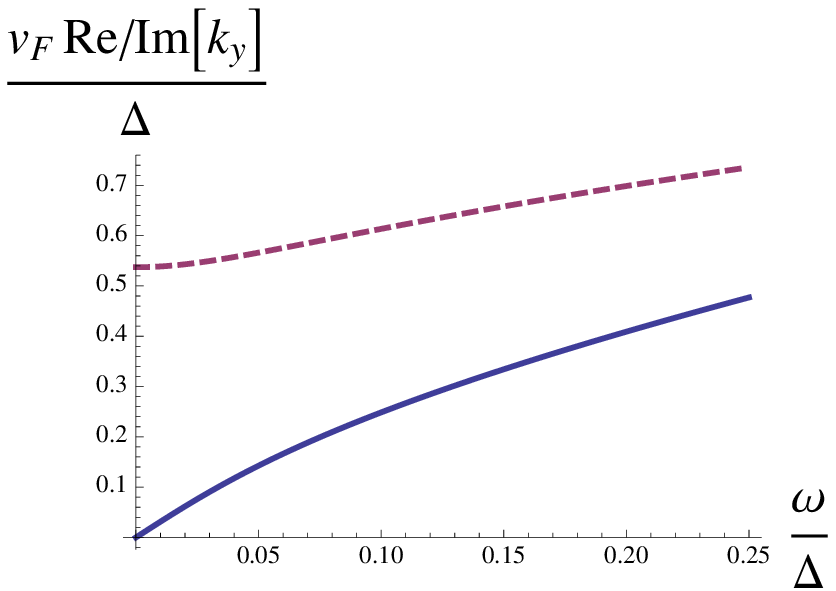}
\caption{Dispersion of the edge states:
  $k_\parallel=\textrm{Re}[\to]/v_F$ (full line) and inverse decay
  length  $L_\parallel^{-1}=\textrm{Im}[\to]/v_F$ (dashed line) versus
  $\omega$. (a) Gapped phase ($\zeta=0.8\,\zeta_{\rm gap}$). (b)
  Gapless phase ($\zeta=1.2\,\zeta_{\rm gap}$).}\label{fig-edge_disp}
\end{figure}

When the disorder is sufficiently weak, $\zeta<\zeta_{\rm gap}$, the
dispersion decribed by \eqref{om} can be separated into two
regimes. At energies below the renormalized gap edge,  $\omega<E_{\rm
  gap}<\Delta$, we find edge state solutions that do not decay, i.e.,
$\textrm{Im}[ \to]=0$. By contrast, at energies $\omega>E_{\rm gap}$,
we find edge states with a finite lifetime,
$\textrm{Im}[\to]\neq0$. Indeed, the finite lifetime of the edge state
is a consequence of the hybridization with the continuum which permits
for the state to decay to extended bulk states. The dispersion and
lifetime of the weak-disorder edge states is shown in
Fig. \ref{fig-edge_disp}a.

\begin{figure}[h]
\includegraphics[width=0.95\linewidth]{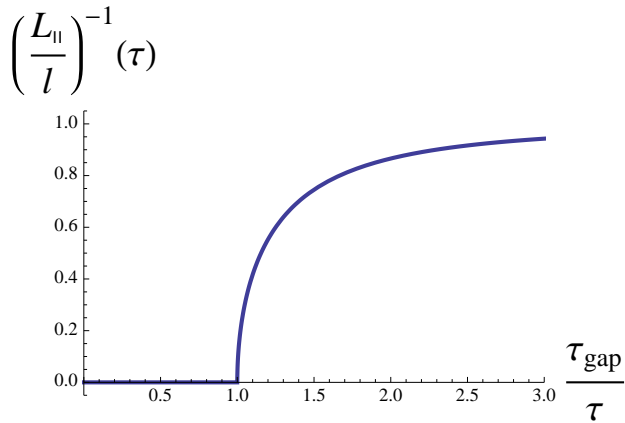}
\caption{Inverse decay length along the edge, $L_\parallel^{-1}$, in
  units of the inverse bulk mean free path, $l^{-1}=1/(v_F\tau)$, as a
  function of disorder. $L_\parallel^{-1}$ is zero in the gapped phase
  and approaches $l^{-1}$ as  $\tau^{-1}\rightarrow\infty$. 
}\label{fig-edge_parallel}
\end{figure}

As the disorder increases, $\zeta>\zeta_{\rm gap}$, the bandgap
vanishes. Now the entire edge state branch is hampered by a finite
lifetime and propagation length, see Fig. \ref{fig-edge_disp}b. The
decay length in the propagation direction $L_{\parallel}$ is nearly
independent of energy, but depends strongly on the disorder
strength. By setting $\omega=\textrm{Re}[\to]=0$, one can obtain
$L_{\parallel}=v_F/\textrm{Im}[\to]$. We find, relying on the
identification of $k_y$ in Eq. (\ref{kxky}),
\be
L_\parallel^{-1}=\frac1{v_F\tau}\sqrt{1-\left(\frac\Delta2\tau\right)^2},
\ee
where $\tau$ is the mean free time as defined in Eq. (\ref{mft}). The
result is shown in Fig. \ref{fig-edge_parallel}. At
$\zeta=\zeta_{\rm gap}$, the decay length diverges,
$L_{\parallel}\rightarrow \infty$. 

Quite interestingly, the survival length of the edge state is closely
related to the topological index calculated in
Sec. \ref{topoind}. Namely,
\be
C=-\frac{2}{\pi}s_z\arctan\frac{L_\parallel\Delta}{2v_F}.\label{CL}
\ee
This gives more ground to the speculation that a non-quantized
topological index will always be associated with a finite lifetime for
the edge states. 

\begin{figure}[h]
(a)\\
\includegraphics[width=0.95\linewidth]{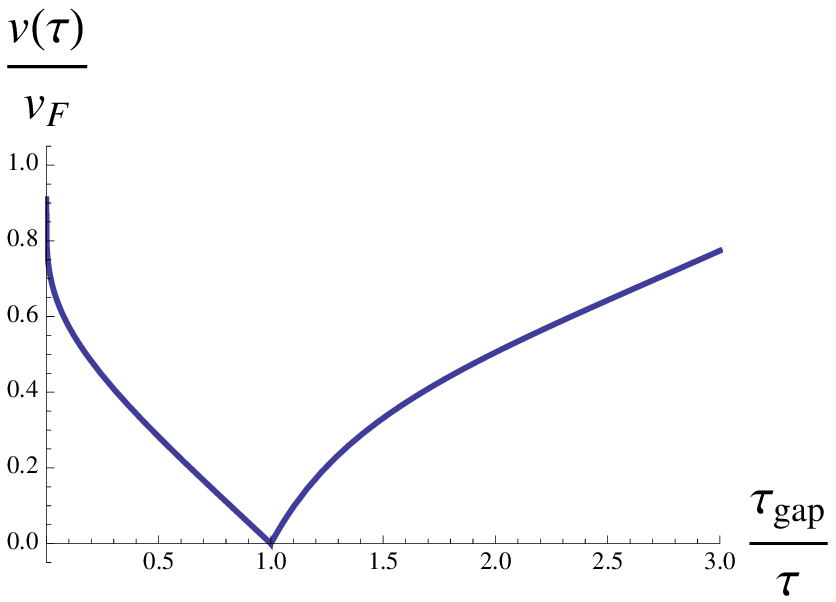}\\(b)\\\includegraphics[width=0.95\linewidth]{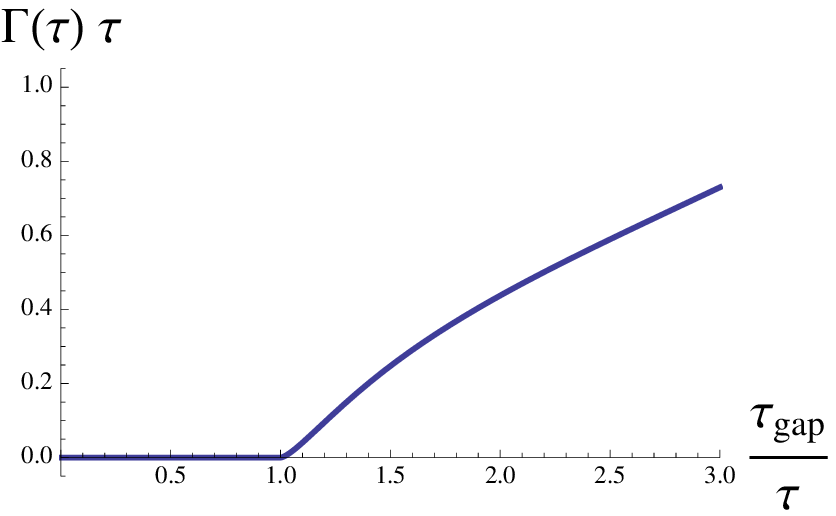}
\caption{Properties of the edge states as a function of the scattering
  rate $\tau^{-1}$. (a) Effective velocity $v$ of the edge states in
  units of the Fermi velocity. (b) Decay rate $\Gamma$ of the edge
  states in units of the scattering rate $\tau^{-1}$. As the velocity
  vanishes at $\zeta_{\rm gap}$, $\Gamma$ increases at a much slower
  rate than $L_\parallel^{-1}$ in the gapless phase, namely
  $\Gamma\propto L_\parallel^{-3}$.}\label{fig-edge}
\end{figure}

Another interesting property of the edge states is the propagation
velocity, $v=v_F(\partial\omega/\partial\tilde\omega)$. In the limit
$\omega\rightarrow0$, straightforward manipulation of Eq. (\ref{om})
leads to 
\be
\frac v{v_F}=\l\{\ba{cc}
1-2\zeta \ln\frac{D(2-\frac v{v_F})}{\Delta} \quad & \zeta<\zeta_c,\\[0.1cm]
\frac{2v_F^2\tau^2L_\parallel^{-2}}{\zeta^{-1}-\left(\frac\Delta2\tau\right)^2}
& \zeta>\zeta_c.
\ea
\rr.
\label{vel}
\ee
The result is depicted in Fig. \ref{fig-edge}a. Approaching the
transition from both sides, we find that the velocity vanishes when
$\zeta=\zeta_{\rm gap}$. In the gapped regime, we identify the
transcendental equation defining $v$ for $\zeta<\zeta_{\rm gap}$ as
reducing to the gaplessness condition \eqref{eq-zg}  upon setting
$v=0$. Indeed, the edge velocity is suppressed throughout the
weak-disorder regime at energies tending to the gap from below,
$\omega\rightarrow E_{\rm gap}^{-}$.   In the gapless regime,
$\zeta>\zeta_{\rm gap}$, the denominator \cite{ft3} in Eq. \eqref{vel}
is regular at $\zeta=\zeta_{\rm gap}$. Thus, since $L_\parallel^{-1}$
vanishes at the transition, so does $v$. 

With the velocity and the decay length, we may now identify the decay
rate of the edge states in the gapless regime as a function of
disorder,
\be
\Gamma=\frac{v}{L_{\parallel}}=
\frac{2v_F^3\tau^2L_\parallel^{-3}}{\zeta^{-1}-\left(\frac\Delta2\tau\right)^2}. 
\ee
Thus the lifetime $\sim\Gamma^{-1}$ diverges at the point where the
gap closes, and follows the third power of the decay length as
disorder further grows (the result is depicted in Fig. \ref{fig-edge}b).

\section{Topological Mott metal \label{GMTI}} 

Topological behavior may also emerge due to electron-electron
interactions rather than a native spin-orbit coupling and band
inversion\cite{Raghu,Franz,Vishwanath,PesinBalents,YBKim}. Under these
circumstances, it is particularly interesting to ask whether a gapless
topological phase can exist when the disorder is strong. The situation is
rather reminiscent of the competition between magnetic impurities and
s-wave superconductivity. It is well known that magnetic impurities
may induce a phase which is s-wave superconducting, albeit gapless
\cite{Gorkov,DeGennes,Maki}. Namely, the disorder suppresses both the
order parameter and the spectral gap, but the two do not vanish
simultaneously. In the gapless phase, while the material has a
superconducting stiffness, its critical current is suppressed. In this
section, we will establish that interacting systems can host a
gapless topological phase as well, albeit in a rather narrow range of
interactions and disorder.

The idea of an interaction-induced topological insulator, i.e., a
topological Mott insulator, was first explored in
Ref. \onlinecite{Raghu} for the case of a honeycomb lattice. It was
shown that a next-nearest neighbor interaction may open a gap which
has opposite sign at the two Dirac nodes of the band structure. Using
the Hartree-Fock approach this can be analyzed self-consistently,
which is the approach we will pursue below. Later work \cite{Franz}
demonstrated that various lattice structures may also be susceptible
to competing symmetry-broken phases, most notably the Kekule phase in
the honeycomb lattice. We will stir clear of the possibility of
competing phases, and concentrate on the range of parameters most
conducive to the formation of topological order. 

In particular, we will consider a honeycomb lattice in 2d with
next-nearest neighbor repulsion of strength $V$. Our starting point is
the tight-binding Hamiltonian,
\begin{eqnarray}
H \eq -t\sum_{\langle i,j\rangle;\sigma}(c_{i\sigma}^\dagger c_{j\sigma}+{\rm h.c.})\\&&+V\sum_{\langle\langle i,j\rangle\rangle}\left(\sum_\sigma c_{i\sigma}^\dagger c_{i\sigma}-1\right)\left(\sum_\sigma c_{j\sigma}^\dagger c_{j\sigma}-1\right).\nn
\end{eqnarray}
To this Hamiltonian, we will add disorder as explained in Eq. (\ref{AG}).

\subsection{Self-consistent Hartree-Fock analysis} 

The symmetry breaking that leads to topological behavior in the
honeycomb lattice occurs on the next-nearest neighbor bonds. The
next-nearest neighbor hopping may acquire an imaginary expectation
value which will then result in the Haldane model for each spin
flavor. A next-nearest neighbor interaction naturally leads to such a
broken symmetry \cite{Raghu,Franz}. The interaction term may be
decoupled using a Hubbard-Stratonovich transformation, yielding the
mean-field Hamiltonian,
\begin{equation}
H = -t\sum_{\langle i,j\rangle;\sigma}(c_{i\sigma}^\dagger c_{j\sigma}\!+\!{\rm h.c.})-2V\!\!\!\!\!\sum_{\langle\langle i,j\rangle\rangle;\sigma,\sigma'}\!\!\!\!\!\chi_{ij}^{\sigma\sigma'}c_{i\sigma}^\dagger c_{j\sigma'},
\end{equation}
together with the self-consistency equation,
\begin{equation}
\chi_{ij}^{\sigma\sigma'}=\langle c_{j\sigma'}^\dagger c_{i\sigma}\rangle.
\end{equation}
After Fourier transformation, the Hamiltonian in the sublattice
(pseudo-spin) basis takes the form
\begin{widetext}
\begin{eqnarray}
H \eq-\sum_{{\bf k},{\bf k}';\sigma,\sigma'}C_{{\bf k}\sigma}^\dagger\begin{pmatrix}\Delta_A^{\sigma\sigma'}({\bf k},{\bf k'})&2t\sum_l e^{-i{\bf ka}_l}\delta_{{\bf k},{\bf k}'}\delta_{\sigma\sigma'}\\2t\sum_le^{i{\bf ka}_l}\delta_{{\bf k},{\bf k}'}\delta_{\sigma\sigma'}
&\Delta_B^{\sigma\sigma'}({\bf k},{\bf k'})\end{pmatrix}C_{{\bf k}'\sigma'},
\end{eqnarray}
\end{widetext}
where $\Delta^{\sigma\sigma'}_{X,ij}=2V\chi^{\sigma\sigma'}_{X,ij}$
and $X=A,\,B$ is the sublattice index. Furthermore, ${\bf a}_l$ (with
$l=1,2,3$) are the lattice vectors of the honeycomb lattice,
connecting the $A$ and $B$ sublattices. 

At low energies, we assume that the interaction opens a topological
insulator gap without breaking time-reversal symmetry, and therefore
$\Delta$ has the form
\begin{eqnarray}
\Delta_X^{\sigma\sigma'}({\bf k},{\bf k'})=\Delta \,{\rm sign}({ k_x})(-1)^{X+\sigma}\delta_{{\bf k},{\bf k}'}\delta_{\sigma\sigma'}
\end{eqnarray}
with $(-1)^{A/B}=\pm 1$. Note that ${\rm sign}(k_x)$ inherits the
valley index, and therefore in a continuum picture will be replaced by
$\tau_z$. Indeed, the low-energy Hamiltonian of the system, upon
linearization of the dispersion near the nodes, reads
\begin{equation}
{\cal H} = v_Fk_x\sigma_x\tau_z-v_Fk_y\sigma_y-\Delta\sigma_z s_z\tau_z,
\label{H-fin}
\end{equation}
which is the same as Eq. \eqref{hh}. However, this Hamiltonian is now supplemented by the self-consistency condition for $\Delta$.
Assuming that the low-energy part of the dispersion is dominant in determining the self-consistent bond order parameter,  
the self-consistency condition takes the form
\begin{eqnarray}
\Delta\eq V\int \frac{d\omega}{2\pi}\int(dk)\;{\rm Tr}\left[G(\omega,{\bf k})\sigma_zs_z\tau_z\right]\label{eq-sc}
\end{eqnarray}
with $G(\omega,{\bf k})$ given by Eq. \eqref{sc1}.

\subsection{Finding the critical disorder and interaction}

Disorder now has two effects. As before, it suppresses the spectral
gap, but, at the same time, it also affects the order parameter: for
any interaction strength which leads to the topological phase, adding
sufficiently strong disorder will destroy that phase. The condition
for the spectral gap to vanish remains the same as before. Namely,
\be
\zeta_{\rm gap}=\frac1{2\ln\left(2\frac D\Delta\right)},
\ee
see Eq. \eqref{eq-zg}. However, $\Delta=\Delta(\zeta_{\rm gap})$ has
now to be determined self-consistently. Substituting the Green
function, Eq. \eqref{sc1}, into the self-consistency condition
\eqref{eq-sc}, we find
\begin{eqnarray}
\Delta\eq 8 V\int\frac{d\omega}{2\pi}\int(dk)\;\left[\frac{\td}{\to^2+v_F^2k^2+\td^2}\right],
\end{eqnarray}
which, upon performing the momentum integration and using Eqs. \eqref{sc2}, reduces to
\begin{eqnarray}
\Delta\eq \frac c{\zeta\Lambda}\int\limits_0^\Lambda d\omega\;(1-\delta),
\label{Ddef}
\end{eqnarray}
with $c=2VD/(\pi^2v_F^2)$, and $\Lambda=D/\Delta$ as defined
earlier. Note that $\delta$ and $\omega$ are related through Eq. \eqref{eq-omegadelta}.

\begin{figure}[h]
\includegraphics[width=0.95\linewidth]{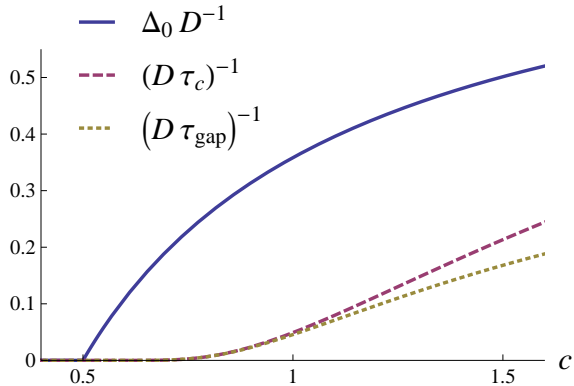}
\caption{Order parameter in the absence of disorder, $\Delta_0/D$, and
  critical scattering rates, $1/(D\tau_c)$ (vanishing of the order parameter) and  $1/(D\tau_{\rm gap})$ (vanishing of the spectral gap), as a function of
  interaction strength, $c$.}\label{fig-selfc}
\end{figure}

Fig. \ref{fig-selfc} shows the order parameter $\Delta_0$ in the absence of disorder. To obtain the critical disorder strength, $\zeta_c$, for which the
topological phase is destroyed,  we  replace the $\omega$-integral
with an integral over $\delta$ and let $\delta/\Lambda\rightarrow
0$. The result is a transcendental equation that connects $c$ and
$\zeta_c$, 
\begin{eqnarray}
1\eq\frac{ce^{-\frac1{2\zeta_c}}}{\zeta_c}\int\limits_{1/2}^1 d\delta\left(2\!-\!\frac1\delta\right)e^{-\frac12\zeta_c^{-1}(\delta^{-1}-2)}.\label{eq-zetac}
\end{eqnarray}
The critical scattering rate $1/\tau_c$ as a function of $c$ is plotted in Fig. \ref{fig-selfc}.

\subsection{Disorder threshold for gaplessness}

\begin{figure}[h]
\includegraphics[width=0.95\linewidth]{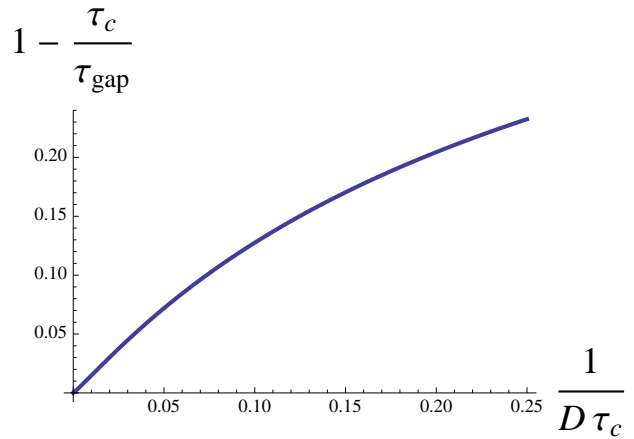}
\caption{Region of gaplessness at $T=0$: $(\tau_c^{-1}-\tau_{\rm gap}^{-1})/\tau_c^{-1}$ as a function of $1/(D\tau_c)$.}\label{fig-zeta_gap-critical}
\end{figure}

To obtain the disorder strength, $\zeta_{\rm gap}$, where the spectral gap vanishes, we need to combine Eqs.  \eqref{eq-sc} and (\ref{Ddef}). 
We find another transcendental equation for $\zeta_{\rm gap}$,
\begin{equation}
1= \frac {ce^{-\frac1{2\zeta_{\rm gap}}}}{\zeta_{\rm gap}}\int\limits_{1/2}^1 d\delta\left(2\!-\!\frac1\delta\right)\sqrt{e^{-\zeta_{\rm gap}^{-1}(\delta^{-1}-2)}\!-\!4\delta^2}.
\end{equation}
The result is plotted in Fig. \ref{fig-selfc}. Furthermore, the width of the
gapless regime is shown in Fig. \ref{fig-zeta_gap-critical}. The larger
$c$ and thus $\Delta_0$, the larger the regime where the order
parameter survives while the gap is closed. In the gapless regime
$\zeta_{\rm gap}<\zeta<\zeta_c$, the above considerations about the
topological index and edge states remain valid, i.e., the system is a
topological Mott metal.

\section{Conclusions}

We have studied the disorder-induced gapless phase of a 2d topological
insulator system. In particular, we have concentrated on (i) the
Kane-Mele model, which is intrinsically topological, as well as (ii)
an example of a Mott topological insulator, namely graphene with
next-nearest neighbor interactions that induce a topological phase.

Our main conclusion is that the disorder-induced gapless system
remains topologically non-trivial. Naively, one would think that once
the gap of a topological insulator is closed, its topological
properties also vanish. Nevertheless, it appears that the gapless
phase retains evidence of its topological origin. Namely, the gapless
phase is characterized by a finite, yet non-quantized
disorder-averaged topological index. This is reminiscent of doped
topological insulators for which
Refs. \onlinecite{HastingsLoring1,HastingsLoring2} showed that the
topological index  changes continuously between 1 and 0 when the
chemical potential is sweeped between the top and bottom of the
valence band.

In addition, we find that the gapless phase still supports helical
edge states, though with a finite lifetime and a renormalized
velocity. This is consistent with the findings of
Ref. \onlinecite{BergmanRefael}, where a bulk parasitic metal
overlapping with edge states was considered in the clean case. There
the individual orbitals of the edges states which overlap in both
momentum and energy with bulk states are split into two states, with
energies above and below the parasitic metallic band. However, these
states leave a \lq ghost\rq\ behind: an edge state with a finite
lifetime, which appears as a resonance in the Green function of the
system, although not being an exact eigenstate of the
Hamiltonian. Furthermore, the finding of a finite lifetime limiting
the manifestations of topological effects may be related to the \lq
parity diffusion mode\rq\ found in Ref. \onlinecite{Shindou1} (see,
e.g., Eqs. (4) and (66) ibid., where the IR cutoff indicates a finite
lifetime). Since that work considered disordered 3d TIs, a direct
comparison, however, is not possible.   

The properties of the edge states/resonances have additional features
worth mentioning. First, it appears that a close link exists between
the decay length of the edge resonances and the non-quantized
topological index, as shown in Eq. (\ref{CL}). Next, disorder strongly
affects the edge state velocity. Weak disorder suppresses the edge
velocity, until it vanishes at the disorder strength where the gap
closes. This is essentially dictated by the fact that the helical edge
branch is confined between the top of the valence band and the bottom
of the conduction band. Interestingly, the velocity picks up as a
function of disorder in the gapless regime, as shown in
Fig. \ref{fig-edge}b. The suppression of the velocity when  $\zeta\sim
\zeta_{\rm gap}$ implies that an electron tunneling into the surface
will remain near the point of tunneling for an extended time. This
effect could be observed through an enhanced zero-bias anomaly due to
a temporary Coulomb blockade
\cite{AltshulerAronov,AltshulerAronovLee,Kane-Fisher}.

If the topological phase is induced by interactions, as considered in
the last part of the paper, the disorder affects both the excitation
gap and the order parameter. In analogy with gapless
superconductivity, we have shown that there is a range of disorder
strengths where a gapless topological phase is realized. In this
regime, the disorder fully suppresses the gap, whereas a finite order
parameter prevails up to a critical disorder strength. Within this
range, the material will have the same properties as the
non-interacting disorder-induced gapless topological phase. 

We acknowledge helpful discussions with Victor Gurarie, Doron Bergman,
and Matt Hastings. This work was funded through an EU-FP7 Marie Curie
IRG (JM), and by DARPA and FENA (GR), the Humboldt foundation, and the
IQIM, an NSF Physics Frontiers Center with support of the Gordon and
Betty Moore Foundation. We also thank the Aspen Center for Physics,
where part of the work was done.

\bibliographystyle{apsrev4-1}
\bibliography{gapless-bib}

\begin{thebibliography}{47}
\bibitem [{ft1()}]{ft1}%
  \BibitemOpen
  \href@noop {} {}\bibinfo {howpublished} {By using the disorder-averaged Green
  functions, we neglect vertex corrections which are expected to be
  irrelevant}\BibitemShut {NoStop}%
\bibitem [{ft2()}]{ft2}%
  \BibitemOpen
  \href@noop {} {}\bibinfo {howpublished} {For interacting systems, Ref.
  \onlinecite{Gurarie} showed that either a pole or zero may appear in the
  Green function}\BibitemShut {NoStop}%
\bibitem [{ft3()}]{ft3}%
  \BibitemOpen
  \href@noop {} {}\bibinfo {howpublished} {Note that the denominator of Eq.
  (\ref {vel}) for $\zeta >\zeta _c$ may vanish. However, this occurs at very
  large values of $\zeta$ where the analysis is far from valid}\BibitemShut
  {NoStop}%
\end{thebibliography}
\end{document}